\documentclass[preprint2]{aastex}

\begin{document}

\title{Direct determination of the spiral pattern rotation speed of the Galaxy}
\author{Wilton S. Dias \footnote{Instituto de Fisica de S\~ao Carlos, Universidade de S\~ao 
Paulo, Caixa Postal 369, S\~ao Carlos 13560-970, SP, Brazil}
\and 
  J.R.D.L\'epine\footnote{ Instituto de Astronomia, Geof\'isica e 
Ci\^encias Atmosf\'ericas, Universidade de S\~ao Paulo, Cidade Universit\'aria, S\~ao Paulo, SP, Brazil; 
E-mail: jacques@iagusp.usp.br}
}

\begin {abstract} 

The rotation velocity of the spiral pattern of the Galaxy is determined by direct
observation of the birthplaces of open clusters of stars in the galactic disk as
a function of their age. Our measurement does not depend on any specific model
of the spiral structure, like the existence of a given number of spiral arms,
or the presence of a bar in the central regions. This study became possible
due to the recent completion of a large database on open clusters by our
group. The birthplaces of the clusters are determined by two methods, one that
assumes that the orbits are circular, and the other  by integrating the orbits
in the Galactic potential for a time equal to the age of the clusters. We selected
in the database a sample of 212 clusters for which proper motions, radial velocities,
distances and ages are available, or of 612 clusters that have ages and distances available.
We tested different assumptions concerning the rotation curve and the radius $R_0$ 
of the solar orbit. Our results confirm that a dominant fraction of the open 
clusters are formed in spiral arms, and that the spiral arms rotate like a rigid
body, as predicted by the classical theory of spiral waves. We find that the
corotation radius $R_c$ is close to the solar galactic orbit ($R_c/R_0 = 1.08 \pm 0.08$).
This proximity has many potentially interesting consequences, like a better preservation
of life on the Earth, and a new understanding of the history of star formation in the solar
neighborhood, and of the evolution of the abundance of elements in the galactic disk.
 
\end{abstract}


\section{Introduction}

According to the classical theory of galactic spiral waves proposed by Lin \& Shu (1964)
and Lin et al. (1969), the spiral arms are restricted to the interval between the inner
and outer Lindblad resonances (ILR and OLR), where the pattern angular velocity
 ${\Omega}_p$ equals $\Omega \mp \kappa/2$, where  $\Omega$ is the angular rotation
velocity of the disk and $\kappa$ is the epicycle frequency. The spiral patern is 
considered to rotate like a rigid disk, while the gas and stars present differential
rotation. The radius where $\Omega = {\Omega}_p$, called corotation radius, is situated
between the ILR and the OLR. Since $\kappa$ is a function of $\Omega$ and of its 
derivative $d\Omega/dr$ only (eg. Binney and Tremaine, 1987), given the rotation curve 
of the Galaxy, the position of the Lindblad resonances depend only on the pattern
rotation speed.  

The radii of those resonances and of corotation in our Galaxy have been a
subject of controversy. Lin and his collaborators situated the corotation at the edge of
the galactic disk, at about 16 kpc from the center. Other authors like Amaral \& L\'epine
(1997) and Mishurov \& Zenina (1999) claimed that the corotation is close to
 the radius of the solar orbit $R_0$. This view is supported by recent studies 
 of metallicity gradients in the galactic disk (Andrievski et al, 2004, Mishurov et al.,
 2002). Yet another group of researchers believes that the spiral pattern rotates so 
 fast that the corotation resonance is situated in the inner part of the Galaxy at
 $r \approx 3-4 \, kpc$ and the OLR  is located close to the Sun (e.g. Weinberg
 1994; Englmaier \& Gerhard 1999; Dehnen 2000, etc.). In parallel with the quasi-stationary
 models just mentioned, that are variations of the model of Lin and Shu, there are
 quite different interpretations of the spiral structure, like those of
 Binney \& Lacey (1988) and of Sellwood \& Binney (2002), who consider that the arms are
  constituted by a series of transient waves with different pattern velocities 
 (and consequently, different corotation radii), and that of  Seiden \& Gerola
 (1979), who argue that the arms are not density waves but
 the result of a stochastic self-propagating star formation process. 
  
It is therefore an important step in the understanding of the spiral structure to 
firmly establish what is the rotation velocity of the spiral pattern in the Galaxy,
as well as to verify if different arms have the same  velocity.  
In the present paper we present a new method to measure the velocity of the 
spiral arms, based on the orbits of a sample of open clusters, which have known
distances, space velocities and ages.

\section{The Open Clusters Catalog}

We make use of the {\it New Catalogue of Optically visible Open Clusters and Candidates} 
 published by Dias et al. (2002) and updated by Dias et al. (2003) {\footnote {Available
 at the web page http://astro.iag.usp.br/{\/~}wilton}}. This catalog updates the previous
 catalogs of Lyng\"a (1987) and of Mermilliod (1995). The present version of the catalog
 contains 1689 objects, of which 599 (35.5\%) have published distances and ages, 612 (36.2\%)
 have published proper motions (most of them determined by our group, Dias et al., 2001, 2003)
  and 234 (13.8\%) have radial velocities.    

\section{Basic assumptions}

The basic assumption is that star formation, and in particular, open cluster formation,
takes place only (or almost only) in spiral arms. This follows from the ideas of
Roberts (1969), Shu et al. (1972), and many others, according to whom the shock waves 
occurring in spiral arms are the triggering mechanism of star formation. There is
observational evidence that this is true in external spiral galaxies, where we see 
the HII regions and massive stars concentrated in spiral arms.
The fact that young open clusters are tracers of the spiral structure has been known
for several decades (eg. Becker \& Fenkart, 1970). The distribution of open clusters
in the Galactic disk is shown in Figure 1, for clusters younger than 7 Myr,
and in Figure 2, for clusters older than 30 Myr . It is clear that in about 20 Myr
the open clusters drift away from the arms and fill the inter-arm regions. 
It will be shown in this work that the structure that has disappeared in Figure 2 can
be retrieved by a proper correction of the cluster positions.  
\begin{figure}
\plotone{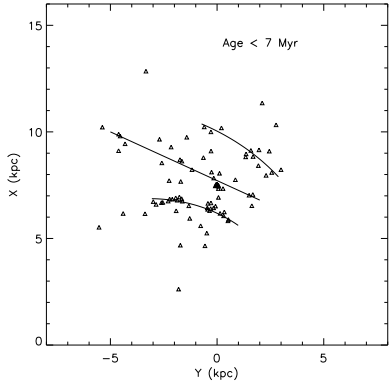}
\caption{The sample of open clusters with age $<$ 7  Myr in the solar 
neighborhood. The Sun is at coordinates (0, 7.5); the Galactic
center at (0,0); distances are in kpc.  }
\label{fig1}
\end{figure}

\begin{figure}
\plotone{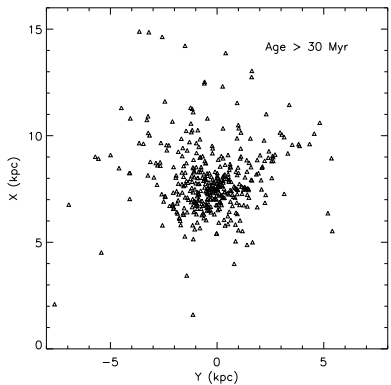}
\caption{The sample of open clusters with age $>$ 30 Myr in the solar 
neighborhood.}
\label{fig2}
\end{figure} 
After their birth, the clusters follow Galactic orbits that can be easily
calculated, since we know the galactic potential. Knowing the present day
space velocity components, we can integrate each orbit backwards for a time
interval equal to the age of the cluster, and find where the clusters were 
born. The birthplace of each cluster indicates the position of a spiral arm at
a past time equal to its age. In principle, it is not difficult to
trace the motion of the arms. Our task is to find the most reliable
way to extract this piece of information from the clusters data.

Before looking for precise methods to determine the corotation radius, we 
call attention to the fact that a direct inspection of the distribution
of the clusters in the galactic plane, gives an approximate value
of that radius, in a way that does not depend on any particular choice of a 
rotation curve or of a set of  galactic parameters. Let us compare Figure 1 
with Figure 3; the difference between the two figures is the range of age of the
clusters. The lines representing the position of the arms in Figure 1 
(age $<$ 7 Myr) are reproduced as guidelines in Figure 3 (12 Myr $<$ age
$<$ 25 Myr). These lines represent approximately the present day position of the arms.
In Figure 3, most of the clusters associated with the Perseus arm (the outer arm in
the figure) are situated to the left of the line. In contrast, most of
the clusters associated with the Sagittarius-Carina arm (the inner arm in 
the figure) are situated to the right of the corresponding line. This means that 
the clusters that were born some 10-20 Myr ago, and since then are rotating
around the galactic center with about the velocity of the rotation curve,
have rotated slower than the spiral pattern at the outer radii, and  
faster than the spiral pattern at the inner radii. This is what we expect to
observe, if the corotation radius is between the two arms, that is, 
close to the Sun.     

\begin{figure}
\plotone{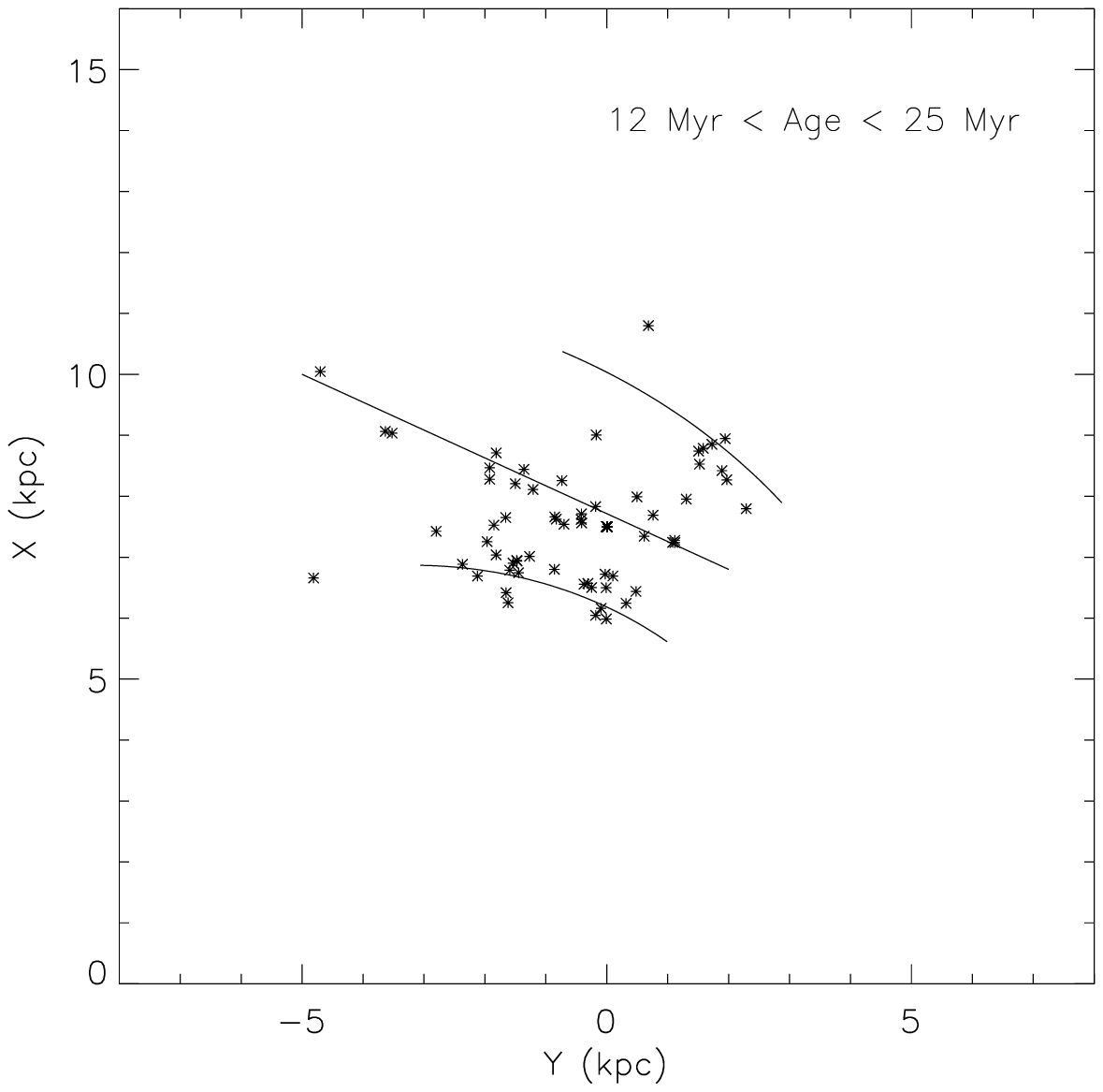}
\caption{The sample of open clusters with 12 Myr $<$ age $<$ 25 Myr in the solar 
neighbourhood. Actual positions are plotted.}
\label{fig3}
\end{figure} 

\subsection{Methods to determine the pattern speed}

We propose two methods to trace back the orbits of the clusters and find their
birthplace, that we shall call "true integration" and "circular rotation", and two
methods to derive the rotation velocity of the arms, based on the birthplaces
of the clusters; we shall call them "direct observation of pattern rotation"
and "reconstruction of the present day arms". We present in this work the best two
combinations of these methods. We next explain the reasons for the use of 
different methods to retrieve the birthplaces. 

In principle, by performing a true integration of the orbit,  using the observed
space velocities as initial conditions, one takes
into account the fact that the orbits are not precisely circular, and one obtains
the best birthplace determinations. However, true integration is limited to the sample
of clusters which have, in the catalog, distance, age, proper motion and radial 
velocity. At the moment, this amounts to a total of 212 objects only. When we further
restrict the sample to some narrow ranges of ages, as required by one of the methods
discussed below, the available number of clusters becomes too small.
Furthermore, one must remember the components of the space velocities
given in the catalog, in particular the radial velocities, are affected by errors,
so that the initial conditions of the orbit integrations have uncertainties. 
These reasons led us to use simple circular rotation as an alternative method.
In this case, the orbits of the clusters are supposed to be circular. The
birthplaces are found by assuming that the clusters moved a distance 
equal to their age multiplied by their velocity, given by the rotation curve.
Therefore, the only parameters needed to recover the birthplaces are
distances and ages; in this case the available sample in our database
amounts to 599 objects.     

\subsubsection{Direct observation of pattern rotation (Method 1)}
The simplest way to proceed, once the birthplaces of the clusters have been obtained, 
is to directly observe  the rotation of the pattern that they form, as illustrated in
Figure 4. We first fitted segments of spirals (indicated by dashed lines)
to the birthplaces of a sample of very young clusters (5-8 Myr, not shown).
The adopted equation of spiral arms is $r =r_i exp(k\theta + \phi_0)$, where $r_i$ is
the initial radius, $\phi_0$  the initial phase angle and $k = tan(i)$ is a constant related
to the inclination $i$ of the arm. We then rotate those spiral segments around the 
Galactic center by varying $\phi_0$, so as to obtain
the best fit of an older sample (9-15 Myr, clusters birthplaces shown as squares, 
fitted solid lines); the  rotation angle in this example is $\alpha$= 10\degr. 
We emphasize that Figure 4 is different from Figure 3 and previous ones, in the sense
that we are plotting the birthplaces, not the present day position of the clusters.
It is important to remark that a same rotation angle fits correctly the different
arms, based on our sample of clusters. 

The variation of the rotation angle with the age of the samples is shown
in Figure 5. The successive age ranges that were used are indicated in the caption.
For each age range, the only parameter used to fit the pattern is the angle
$\alpha$, like in Figure 4. In order to maximize the number of clusters in 
each sample, we allowed a small overlap of the age ranges, and we adopted
the birthplaces obtained by circular rotation of the clusters, as already
explained. 

The procedure to find the best fit parameter $\alpha$ for each age 
range was the following. For each of the 3 arms, we selected as belonging to the
arm the clusters that are at a distance smaller than 0.5 kpc from it. For
these clusters, we computed the sum of the components of the distances to the arm,
in the Y direction (defined here as the  horizontal direction in the figure,
which is about the direction of rotation of the Galaxy, for objects close to
the Sun). When this sum is zero, this means that there are about the same
number of clusters to the right and to the left of the arm, which gives 
the best fit.

The birthplaces depend obviously on the adopted rotation curve. As later 
discussed, we tested different rotation curves and values of R$_0$.
In Figure 5 we compare the results for two flat rotation curves, with 
V$_0$= 170 kms$^{-1}$ and V$_0$=190 kms$^{-1}$, for R$_0$ = 7.5 kpc.  The slopes
of the fitted lines are respectively 1.16 \degr/Myr and 1.33 \degr/Myr.
It is an expected result that when we adopt a larger rotation velocity,
the birthplaces are found more distant from the solar neighborhood, and a
larger rotation velocity of the pattern is derived. Note that 1\degr/Myr
is equivalent to 17 kms$^{-1}$kpc$^{-1}$, the derived corotation radii
(= V$_0/ \Omega_p$) are 8.6 and 8.1 kpc respectively in this example

\begin{figure}
\plotone{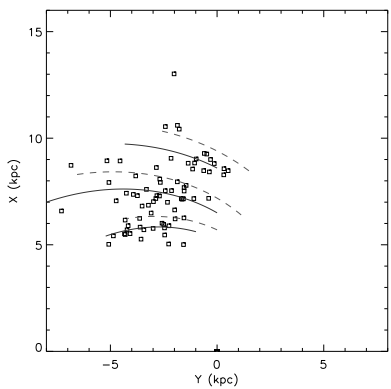}
\caption{Birthplaces of the clusters with ages in the range 9-15 Myr (average 11.6 Myr),
 in the galactic plane. The dashed lines were fitted to a younger sample, not shown,
 with ages in the range 5-8 Myr (average 6.3 Myr); the solid lines are the same arms
 of the dashed line, rotated by 10 \degr around the galactic center. This angle is
 the best fit to the sample displayed.}   
\label{fig4}
\end{figure}

\begin{figure}
\plotone{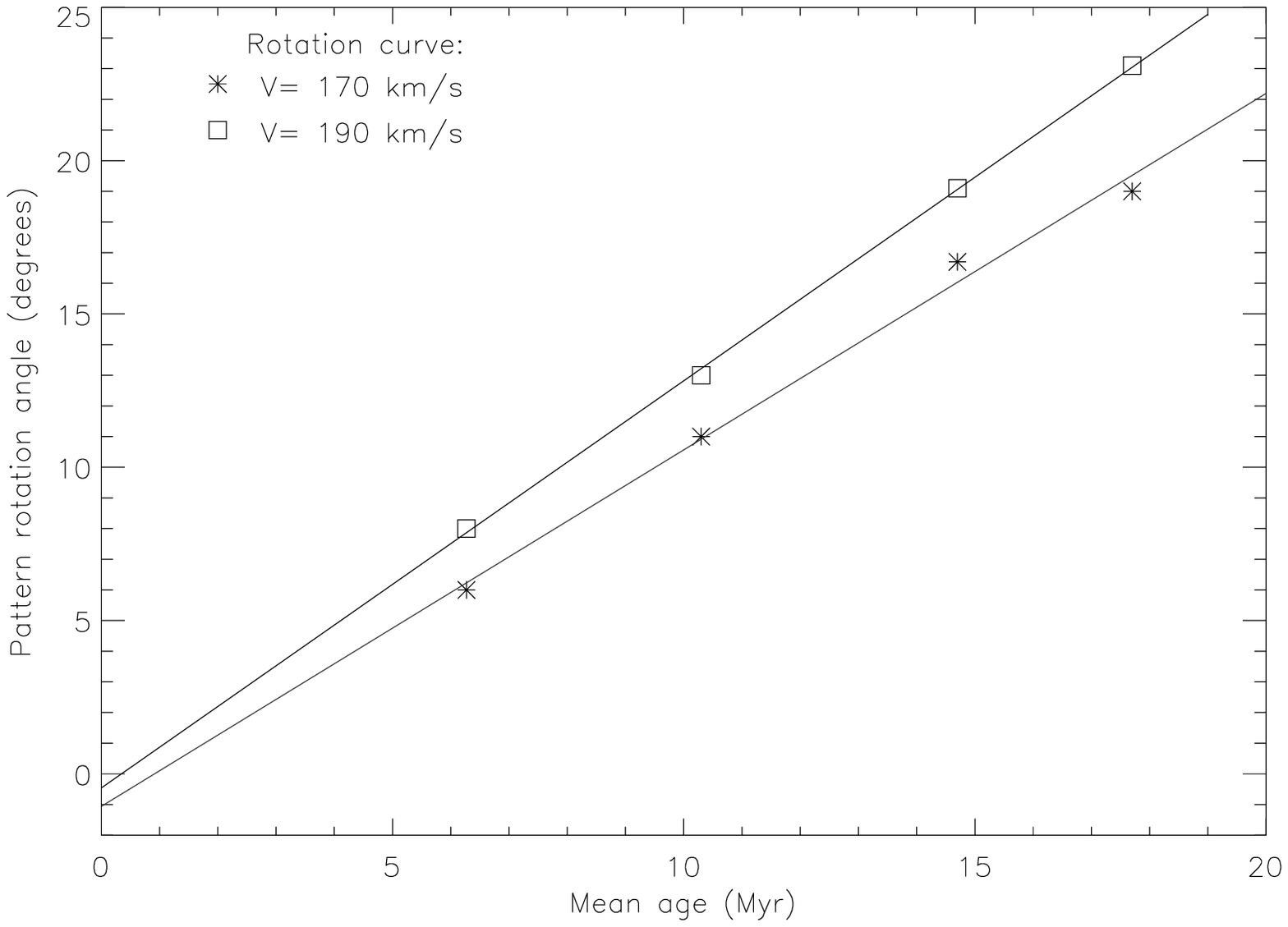}
\caption{Rotation angle required to fit the birthplaces of  samples of clusters 
of different ages with a spiral pattern that represents the present day structure
(Method 1). In this example, the birthplaces were determined with  two flat
(constant velocity)  rotation curves, with V$_0$= 170 kms$^{-1}$ and V$_0$= 190 kms$^{-1}$.
The age range of the samples, in Myr, were 0-8 (average 6.27), 7-14 (average 10.3),
12-18 (average 14.7), 15-22 (average 17.7)}   
\label{fig5}
\end{figure} 

\subsubsection{Reconstruction of spiral arms (Method 2)}

Although it is usually accepted that young galactic objects have
near-circular orbits, it is important to check if the hypothesis of circular
motion can be responsible for systematic errors in the determination
of the pattern rotation speed. For instance, it is expected that the 
shock waves in the spiral perturbation potential produces a braking
the molecular clouds (and consequently, of the recently formed stars)
at galactic radii smaller than the corotation radius, and an acceleration
of the molecular clouds at larger radii. It is therefore necessary to perform 
experiments with exact integration of the orbits, in order to take into
account the initial velocity perturbation.

The method previously described to derive the pattern rotation velocity
requires a sufficiently large number of clusters in each age bin. The second
 method that we propose here does not separate the clusters into age bins, and  
makes use of a sample with a wide range of age. This is  
convenient to work with the small sample of clusters for which
it is possible to perform exact integration of the orbits.

In this method, as a first step, we find the birthplace of each cluster,
integrating backwards their orbits for time intervals $T$ equal to their age, starting
from the present day initial conditions (positions and space velocities).
The method of integration is discussed later. The birthplace of a cluster is supposed 
to represent a point of a spiral arm, a time $T$ ago. If we rotate forward this point
an angle $\Omega_p * T$ around the galactic center, we obtain a point situated on
the present day position of the arm. In this way, points corresponding to very different ages
can be used to trace the present day arms. The unknown is $\Omega_p$. The best estimate 
of $\Omega_p$ is the one that gathers the maximum number of points close to the present day position 
of the arms. In practice, the best value of $\Omega_p$ is found in an interactive mode,
using a technique similar to that of the previous method. We select the points that are
within 0.5 kpc of any of the 3 arms (to avoid mixing of arms), and we minimize the rms distance 
to the arms, varying $\Omega_p$. However, we do not know a priori the 
present day position of the arms, because there are almost no cluster younger 
than 5 Myr. Even if we selected a young sample, we would need to correct the observed positions,
using $\Omega_p$. Therefore, we use an interactive method: for each adopted $\Omega_p$,
we allow minor variations in the parameters describing the arms, in order to
minimize the rms distance of the points to them. Next we compare the rms distances 
obtained with different  values of $\Omega_p$. 
We illustrate in Figure 6 the results obtained with this method, using a CO-based
rotation curve (see below). The adjusted points were  from the sample
of clusters in the range 10-30 Myr. The same result is obtained if we use
another range, like 10-50 Myr. However, the absolute errors on the age tend
to increase with the age, and it is preferable not to exceed 30 Myr.
The derived value of $\Omega_p$ is 25 $\pm$1 kms$^{-1}$kpc$^{-1}$, which correspond 
to a corotation radius of 7.6 $\pm$ 0.3 kpc, for $R_0$=7.5 kpc and $V_0$= 190 kms$^{-1}$.

\begin{figure}
\plotone{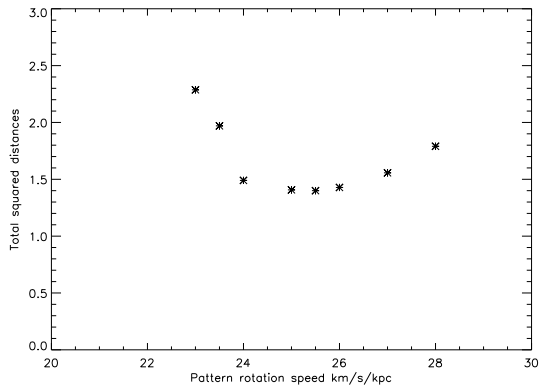}
\caption{Procedure to find the best estimate of pattern rotation velocity
$\Omega_p$ (Method 2). The best value (25 kms$^{-1}$kpc$^{-1}$)is given
by the minimum for sum of the squared distances from the points to the fitted arms. 
This example corresponds to $R_0$=7.5 kpc and $V_0$= 190 kms$^{-1}$.}   
\label{fig6}
\end{figure}

\subsubsection{Rotation curves and details of orbit integration} 

Most of the open clusters in the catalog of Dias et al. are situated at galactic radii
in the range R$_0 \pm 2$ kpc, so that our calculations only make use the
portion of the rotation curve situated in this range. Since the rotation
curve of the Galaxy is known to be relatively flat close to the Sun, a linear
approximation of it is sufficient, in this interval. The only important parameters
are the rotation velocity of the LSR, V$_0$, and the slope of the rotation
curve, (dV/dR)$_{R0}$. These parameters are linked together and with the 
value of R$_0$ through the Oort's constants A and B, which are determined 
by observations ( V$_0$/R$_0$  = A - B,  (dV/dR)$_{R0}$ = -A-B ).
The R$_0$ value 8.5 kpc, recommended by the International Astronomical Union,
is often used in the literature with V$_0$ = 220 kms$^{-1}$, which corresponds
to A-B = 26 kms$^{-1}$kpc$^{-1}$ (eg. Binney \& Tremaine, 1987). This value of A-B seems 
to be well established, and is confirmed by recent observations of a 
different nature (Kalirai et al., 2004) which give 25.3 $\pm$ 2.6 kms$^{-1}$kpc$^{-1}$.
Recent works often adopts R$_0$ = 7.5 kpc (Racine \& Harris, 1989, Reid, 1993,
and many others). The shorter scale is supported by VLBI observations of
H$_2$O masers associated with the Galactic center.  Keeping the same value
for A-B, the corresponding V$_0$ would be 195 kms$^{-1}$. We note that Olling \& 
Dehnen (2003) argue that the rotation curve is flat (A = -B), but usually,
the rotation curve is considered to be decreasing near the Sun, with -A-B 
about -3 kms$^{-1}$kpc$^{-1}$ (Binney \& Tremaine, 1987).

We calculated $\Omega_p$ for two values of R$_0$, 7.5 and 
8.5 kpc, each combined with different values of V$_0$ (170, 190 and 210 kms$^{-1}$ for
R$_0$ = 7.5 kpc, and 180, 200 and 220 kms$^{-1}$ for R$_0$ = 8.5 kpc), considering 
both flat curves and curves similar to that of Clemens (1985). The original
curve obtained by Clemens assumed R$_0$ = 8.5 kpc and V$_0$ = 220 kms$^{-1}$.
We used the same observational CO data of Clemens to reconstruct the curve for different
values of R$_0$ and V$_0$. In each case we fitted the curve with a simple
analytical expression, in order to use it in the calculations. An example
of these CO-based curves, for  R$_0$= 7.5 kpc and V$_0$= 190 kms$^{-1}$, is shown
in Figure 7. The fitted curve has a linear behavior
close to the Sun, with a slope about -3.3 kms$^{-1}$kpc$^{-1}$. For this curve in particular,
both V$_0$/R$_0$ and (dV/dR)$_{R0}$ are in agreement with the best estimate
of Oort's constants.

To perform numerical integration of the orbits, the observed radial velocities and
proper motions were corrected for differential rotation before being converted to
U, V, W velocities in the galactic local system. The corrections in radial velocity, 
and proper motions in longitude and latitude directions ($\mu_l$ cos(b) and $\mu_b$)
 are respectively (eg Carraro \& Chiosi ,1994): $C_{rv}= A d cos2(b) sin(2l)$ ;
  $C_{pml} = (A cos(2l) +B) cos(b)/ 4.74$ ; $C_{pmb} =  A sin(2l) sin (2b)/ 2×4.74$ ,
 where $l,b$  are the galactic coordinates, $d$ the distance,
 $A, B$ the Oort's constants. For each galactic rotation curve that we used, we first 
made the proper motion correction using the corresponding values of A and B. Another 
correction is the transformation of the velocities to the Local Standard of Rest; we
 assumed that the Sun has a velocity 22 kms$^{-1}$ in the direction $l=61.4\degr,
 b= 20.4\degr$ (Abad et al., 2003). 

\begin{figure}
\plotone{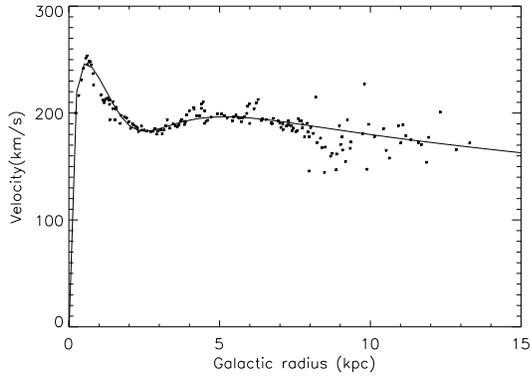}
\caption{CO-based rotation curve of the Galaxy, with Clemens (1985) data 
corrected for R$_0$=7.5 kpc  and V$_0$ =190 kms$^{-1}$. The curve is fitted by 
the expression $V= 228~exp(-r/50 -(3.6/r)^2) + 350~exp(-r/3.25 - 0.1/r)$}   
\label{fig7}
\end{figure} 

\section{Results and conclusions}

The galactic  corotation radii $R_c$ derived in this work are summarized in
Table 1. Typical errors on individual determinations are about $\pm$ 0.3 kpc.
The values of $R_c$ that we obtain with different methods and 
different sub-samples of open clusters are quite consistent. $R_c$ is slightly larger
 than $R_0$ in almost all determinations;  $R_c/R_0$ is situated in the interval 1.02 to
1.14, the best estimate being $R_c/R_0$ = 1.08 $\pm$0.08.  $R_c$ is roughly proportional to
the adopted value of $R_0$, and depends very little on $V_0$. Within the precision of our
method and the limitations of our sample, the three main arms seen in the solar neighborhood
present the same rotation velocity. Our results strongly favor the idea that the spiral
pattern rotates like a rigid body. The pattern rotation speed is 25 kms$^{-1}$kpc$^{-1}$,
which situates the ILR and OLR at 2.5 and 12 kpc, respectively, for $R_0$ = 7.5 kpc and
$V_0$ = 190 kms-1. Considering that the Galaxy has an important 4-arms mode, the
corresponding ILR and OLR ($\Omega \pm \kappa/4$) are at about 4 kpc and 10 kpc. For
almost every open cluster, we can retrieve a birthplace which coincides with the position
of a spiral arm at the epoch of its birth. This observation, which is not feasible 
in external galaxies, gives robust support to the view that spiral arms are the triggering
mechanism of star formation.

\begin{deluxetable}{cccccccc}
\tabletypesize{\scriptsize} \tablecaption{Co-rotation radius obtained 
with different rotation curves. \label{tbl-1}} \tablewidth{0pt}
\tablehead{\colhead{Curve} & \colhead{ Method} & \colhead{R$_0$ (kpc)} &
\colhead{V$_0$ (km/s)}& \colhead{R$_C$ (kpc)}} \startdata
Flat & circ. rot. & 7.5    &  170 & 8.6  \\     
Flat & circ. rot. & 7.5    &  190 & 8.1  \\  
Flat & circ. rot. & 7.5    &  210 & 8.0  \\      
Flat & circ. rot. & 8.5    &  190 & 9.4  \\       
Flat & circ. rot. & 8.5    &  210 & 9.7  \\
CO based & circ rot. & 7.5 & 170 & 7.5    \\
CO based & circ rot. & 7.5 & 190 & 7.9    \\
Flat & true integr. & 8.5    &  170 & 8.1  \\   
Flat & true integr. & 8.5    &  190 & 8.6  \\     
CO based & true integr. & 7.5    &  170 & 8.1  \\   
CO based & true integr. & 7.5    &  190 & 7.6  \\     

\enddata
\end{deluxetable}
 The proximity of the Sun to the corotation radius means that it has a small
 velocity with respect to the spiral arms, and that long periods of time elapse between 
successive crossings of the spiral arms. The crossing of spiral arms is a probable
explanation for the peaks in the history of star formation in the solar neighborhood
(Rocha-Pinto et al., 2000, de la Fuente Marcos \& de la Fuente Marcos, 2004).
Furthermore, the encounters with spiral arms, with the 
larger probability of nearby supernovae explosions and of gravitational perturbation
of the Oort cloud, making more objects like comets to fall towards the inner solar system,
could be associated with events of mass extinction of terrestrial (or extra-terrestrial)
life (Leitch \& Vasisht, 1998). Finally, the corotation radius is often considered to
be associated with a minimum of star formation. If the star formation rate (SFR) is 
related to the rate at which the interstellar gas is introduced into the arms, that
can be viewed as gas-to-star transformation machines, a recipe that has been proposed
is  SFR $\propto \mid \Omega - \Omega_p \mid$  (Mishurov et al., 2002). 
The minimum in the star formation rate close to the Sun is a condition that favors the
survival of life on the Earth. Moreover, a radius with a minimum of star formation rate
 must correspond to a minimum of metallicity, since the enrichment of the interstellar
 medium in metals is related to the death of short-lived massive stars. These concepts
 are essential to understand the bimodal gradient of abundance of different
 elements in the disk (Andrievsky et al., 2004).

\acknowledgments{The work was supported in part by the Sao Paulo State
agency FAPESP}

\end{document}